\documentclass[aps,preprint,superscriptaddress,nofootinbib]{revtex4-1} 
\usepackage{color}
\usepackage{amsfonts}
\usepackage{amsmath}
\usepackage{amssymb}
\usepackage{graphicx}
\usepackage{epstopdf}
\usepackage{epsfig}
\newcommand{\be}{\begin{equation}}
\newcommand{\ee}{\end{equation}}
\newcommand{\bea}{\begin{eqnarray}}
\newcommand{\eea}{\end{eqnarray}}
\newcommand{\ba}{\begin{array}}
\newcommand{\ea}{\end{array}}

\def \nn {\nonumber}
\newcommand{\eq}[1]{(\ref{#1})}

\newcommand{\tr}{\mbox{tr}}

\newcommand{\p}{\partial}

\newcommand{\s}{\sigma}
\newcommand{\e}{\epsilon}

\newcommand{\om}{\omega}

\newcommand{\cL}{{\cal{L}}}

\newcommand{\beq}{\begin{eqnarray}}
\newcommand{\eeq}{\end{eqnarray}}
\newcommand{\bes}{\begin{subequations}}
\newcommand{\ees}{\end{subequations}}

\def \nn {\nonumber}
\def \p {\partial}
\def \th {\theta}
\def \G {\Gamma}
\def \Ld {\Lambda}

\def \e {\epsilon}
\def \om {\omega}

\def \a {\alpha}
\def \b {\beta}

\def \g {\gamma}
\def \d {\delta}
\def \k {\kappa}

\def \et {\eta}
\def \ps {\psi}
\def \ph {\phi}
\def \m {\mu}
\def \n {\nu}
\def \l {\lambda}
\def \s {\sigma}
\def \S {\Sigma}
\def \t {\tau}
\def \nb {\nabla}
\def \tr {\mbox {tr}}

\def \ve {\sqrt{-g}}
\def \sk {\textsc{k}}

\begin{document}


\title{Relative Entropy and Torsion Coupling}

\author{Bo Ning} \email{ningbbo@gmail.com}
\affiliation{Center for Theoretical Physics, College of Physical Science and Technology,
Sichuan University, Chengdu, 610064, PR China}
\author{Feng-Li Lin} \email{fengli.lin@gmail.com}
\affiliation{Department of Physics, National Taiwan Normal University, Taipei, 116, Taiwan}

 \date{\today\\
 \vspace{1.6in}}
\begin{abstract}
  
    We evaluate the relative entropy on a ball region near the UV fixed point of  a holographic conformal field theory deformed by a fermionic operator of nonzero vacuum expectation value. The positivity of the relative entropy considered here is implied by the expected monotonicity of decrease of quantum entanglement under RG flow. The calculations are done in the perturbative framework of Einstein-Cartan gravity in four-dimensional asymptotic anti-de Sitter space with a postulated standard bilinear coupling between axial fermion current and torsion. By requiring positivity of relative entropy, our result yields a constraint on axial current-torsion coupling, fermion mass and equation of state.

\end{abstract}
\maketitle
\tableofcontents

\setcounter{footnote}{0}

\section{Introduction}

    The AdS/CFT correspondence relates the bulk gravity to the boundary dual CFT in a nontrivial way. It has inspired numerous studies and shed light on our understanding for both quantum gravity and strongly coupled CFT in the past twenty years. In particular, the advance of the Ryu-Takayanagi formula \cite{Ryu:2006bv,Ryu:2006ef} ten year ago brings the information perspective into the studies of quantum gravity and CFT, and thus deepens our insight beyond the conventional scope. Many entanglement-related quantities in quantum information of CFT then have the dual geometrical realizations, see \cite{VanRaamsdonk:2016exw} for a recent review on this aspect.
            
      An important concept in the information theory is the relative entropy \cite{relativeE-1,relativeE-2,relativeE-3,Casini:2008cr,Wall:2010cj,Wall:2011hj} which measures the ``distance" between two states. Holographically, different CFT states are dual to different bulk geometries so that the relative entropy can be viewed as some kind of bulk geometric measure. 
For the nearby states/geometries the relative entropy is zero at the first order. This fact is also dubbed as the first law of entanglement thermodynamics \cite{Bhattacharya:2012mi}, which relates the change of the entanglement entropy to the change of the modular Hamiltonian. Later the first law is shown to be holographically dual to the linearized Einstein equation of the bulk gravity \cite{Blanco:2013joa,Wong:2013gua,Faulkner:2013ica,Nozaki:2013vta,Bhattacharya:2013bna,Guo:2013aca,Swingle:2014uza}.  Using the first law, it is further shown recently that the entanglement entropy of a ball region obeys Laplacian equation of the auxiliary de Sitter space, i.e., kinematic space of geodesics \cite{deBoer:2015kda,Czech:2016xec,deBoer:2016pqk}.

    As a measure of ``distance" between states, the relative entropy is positive definite by its mathematical definition. The violation of such positivity would imply pathology of the underlying quantum field theories, such as the violation of unitarity.  One can then take this property as diagnostic for pathological coupling space, i.e., swampland of the holographic CFTs.  Due to the convexity of the relative entropy, the first order result is always zero so that we need to evaluate the relative entropy at least up to the second order to achieve the aforementioned diagnosis. This was firstly initiated in \cite{Blanco:2013joa}, which showed that the positivity of the relative entropy is consistent with the Breitenlohner-Freedman (BF) bound on the mass of bulk scalar. Recently, there are intensive studies on the second order results \cite{Lin:2014hva,Lashkari:2014kda,Lashkari:2015hha,Jafferis:2015del,Lashkari:2016idm}.  Especially, a key result shown in \cite{Lashkari:2014kda,Lashkari:2015hha,Lashkari:2016idm} is that the positivity of the relative entropy of a ball region at the second order is equivalent to the positive quasi-local energy condition of the bulk gravity. This implies that the quantum information inequality of a holographic CFT is guaranteed by the stability of the bulk Einstein gravity.  Another interesting result as shown in \cite{Jafferis:2015del,Jafferis:2014lza} suggests that the relative entropy of the holographic CFT states equals to the relative entropy of the associated bulk quantum states, which is relevant to the entanglement wedge reconstruction of the bulk operators \cite{Dong:2016eik,Almheiri:2014lwa}. 
    
      Along the above line of research, in this paper we will exploit the positivity of the relative entropy to constrain the holographic CFT, which is dual to the torsion gravity.   According to AdS/CFT correspondence, the torsion field should be dual to a new CFT operator which is not considered before in the literatures as far as we know. Though a CFT is mainly dictated by the conformal symmetry, its spectrum and the OPE coefficients are severely constrained by unitarity, bootstrap conditions \cite{ElShowk:2012ht} and the information inequalities such as the aforementioned positivity of relative entropy. Therefore, it is interesting to explore such kind of constraints on the additional new torsion operator, which should also constrain the bulk torsion gravity.

      Torsion as the antisymmetric part of an asymmetric affine connection was firstly discussed by Eddington \cite{Eddington} and then formulated by Cartan \cite{Cartan}.  One may introduce torsion when there are microscopic fermion matters coupled to gravity through spin connection because there is a possibility that the fermions will source torsion. As the gauginos and gravitinos are the necessary ingredients of the supergravities, it is natural to introduce torsion and to discuss its dynamical role in this context.  On the other hand, there seems no consideration of torsion gravity in the context of holographic principle because it is usually not necessary to take the backreaction of the bulk fermion matters into account. This is not the case in this paper when we consider the holographic relative entropy by evaluating the backreacted bulk metric due to the bulk fermion condensate.  We hope that our holographic results will motivate future works from CFT analysis on the similar constraint due to the torsion operator.

      To be specific, we start with the canonical formulation of torsion gravity, the so-called Einstein-Cartan gravity \cite{Hehl:1976kj,Shapiro:2001rz,Cai:2015emx} in which the metric covariantly couples to the torsion and non-torsion parts of the affine connection with the same coupling strength. We call this the canonical coupling to torsion. When introducing the bulk fermion dual to a fermionic CFT operator, one is free to postulate a bilinear interaction between the fermionic axial current and the torsion with arbitrary coupling constant $\eta_t$, i.e., 
\be\label{LpsiK}
\cL_{\psi K} = \frac{\eta_t}{4!} \ve \, \epsilon_{\m\n\rho\sigma} \,{\bar \ps} \g^{\m}  \g_5 \ps \, K^{\n \rho\sigma}
\ee
where $K^{\n \rho\sigma}$ is the contorsion tensor.  If $\eta_t=1$, $\cL_{\psi K}$ is the canonical coupling of the fermion to torsion as conventionally chosen in \cite{Hehl:1976kj,Shapiro:2001rz}.  Now we can diagnose if such a coupling is favored by the positivity of the relative entropy.        
         
    We will evaluate the relative entropy in the perturbative framework of Einstein-Cartan gravity  in four-dimensional asymptotically anti-de Sitter space (AdS$_4$) by solving the backreacted metric due to the zero modes of the bulk fermion.  However, after some calculation we realize that the relative entropy is zero at all orders of gravitational coupling for the fermionic excited states. This is because the bulk stress tensor of the bulk fermions vanishes for such holographic states, for which we turn on either normalizable or non-normalizable mode but not both.  It then seems that there is no constraint from the positivity of relative entropy. This is not the case if we turn on both the normalizable and non-normalizable modes of the bulk fermion, the nonzero bulk stress tensor will then backreact to bulk geometry to yield nonzero relative entropy at the second order. For simplicity, we will only consider turning on the bulk fermionic zero modes. Holographically, this is dual to deforming the holographic CFT by such a coupling term 
\be\label{couplingterm}
\delta \lambda \int d^d x \; \mathcal{O}_{\Delta}(x)
\ee
and with
\be\label{Ovev}
\langle \int d^d x \; \mathcal{O}_{\Delta}(x) \rangle \ne 0
\ee
where $\mathcal{O}_{\Delta}$ is some single-trace fermion operator of conformal dimension $\Delta$, and $\delta \lambda$ is a spinor characterizes the amount of deformation\footnote{So that the overall coupling term \eq{couplingterm} is a scalar as it should be, see \eq{singletrace} for an explicit expression.}.   As expected, the holographic stress tensor is no longer traceless for the deformed fermion state.

   Due to the deformation \eq{couplingterm} and \eq{Ovev} the relative entropy considered in this paper is obtained by comparing some nearby UV states under the RG flow \footnote{One may raise the issue about comparing the states in different theories if their Hilbert spaces are not compatible. To lift such kind of concerns we will assume our comparison is restricted to the nearby region of the UV fixed point, i.e., by requiring $\delta \lambda$ in \eq{couplingterm} to be very small.}. This is in contrast to the previous discussions \cite{Lin:2014hva,Lashkari:2014kda,Lashkari:2015hha,Jafferis:2015del,Lashkari:2016idm} in which they compared the states in the same theory.  
In fact, the CFT calculation of such relative entropy for deformed CFT had been considered \cite{Faulkner:2014jva}, in which it was shown that the relative entropy (or change of entanglement entropy at the second order)  equals  the  integration of bulk energy density of the matter field over the entanglement wedge.  In this paper we will instead evaluate the relative entropy holographically for the deformation caused by fermionic operators, and then require the positivity of the relative entropy to constrain the bilinear coupling $\eta_t$ in \eq{LpsiK} between axial fermion current and torsion. Note that this coupling cannot be fixed by the dynamical symmetry of the holographic CFT. However, we find that the positivity of relative entropy impose some constraint such a coupling. Our results give the example of constraining the dual gravity with novel coupling by the information inequality of the holographic deformed CFT.

\section{Review and Summary}

   In the this section, we will briefly review the basics of relative entropy and its holographic realization. This will form the setup for our consideration in this paper. 
   
\subsection{Relative Entropy and Holographic Consideration}

   The relative entropy \cite{relativeE-1,relativeE-2,relativeE-3} is a ``distance" measure on the (quantum) state space. For two state $\rho$ and $\sigma$, the relative entropy is defined as
\be
S(\rho||\sigma):=\tr(\rho \log \rho)-\tr(\rho \log\sigma)\;.
\ee
As a ``distance" measure, the relative entropy is non-negative, i.e.,
\be
S(\rho||\sigma)\ge 0
\ee
in which the equality holds if and only if $\rho=\sigma$.   Moreover, it is also monotonic, i.e., 
\be
S(\rho_A||\sigma_A)\le S(\rho_B||\sigma_B)
\ee 
if $A\subset B$, where $\rho_A$ and $\sigma_A$ ($\rho_B$ and $\sigma_B$)  are the reduced density matrices for the subsystem $A$ ($B$). 

   When considering the thermal states at temperature $T$, i.e., $\sigma={e^{-H/T} \over \tr e^{- H/T}}$ where $H$ is the Hamiltonian, the relative entropy can be expressed as
\be
S(\rho||\sigma)={1\over T} \Big( F(\rho)- F(\sigma) \Big)
\ee
where the free energy $F(\rho)$ is given by
\be
F(\rho)=\tr (\rho H)- T S(\rho) \;.
\ee
The positivity of the relative entropy around the thermal states was used to prove the Bekenstein bound \cite{Casini:2008cr} and the generalized second law \cite{Wall:2010cj,Wall:2011hj}.     
    
  On the other hand, for a quantum state one can express the reduced density matrix $\sigma_A$ on the region $A$ in terms of the modular Hamiltonian $H_A$ as follows:
\be
\sigma_A={e^{-  H_A} \over \tr e^{-  H_A}}\;,
\ee
then the relative entropy can be rewritten as
\be\label{relativeE-2}
S(\rho_A||\sigma_A)=\Delta \langle H_A \rangle -\Delta S_A
\ee
where
\bea
\Delta \langle H_A \rangle &:=& -\tr(\rho_A \log \sigma_A) + \tr(\sigma_A\log \sigma_A)\;, 
\\
\Delta S_A &:=& -\tr(\rho_A \log \rho_A) + \tr(\sigma_A \log \sigma_A)\;. 
\eea
Note that the entanglement entropy for state $\sigma$ on the region $A$ is defined by 
\be
S_A(\sigma):=-\tr(\sigma_A\log \sigma_A)
\ee
so that $\Delta S_A$ is the difference of entanglement entropy of the region $A$  between states $\rho$ and $\sigma$.   Via \eq{relativeE-2} the positivity of relative entropy yields
\be
\Delta \langle H_A \rangle \ge \Delta S_A\;.
\ee

     The modular Hamiltonian $H_A$ is in general nonlocal and unknown. However, for CFT$_d$ with the interested region $A$ to be a disk of radius $R_A$, it has a closed form as follows \cite{Casini:2011kv}:
\be\label{modular H-ball}
H_A=2\pi \int_{|x|< R_A} d^{d-1}x  \; {R_A^2 - r^2 \over 2 R_A} \; T_{tt}(\vec{x})
\ee
where $T_{tt}$ is the energy density operator of CFT.  When evaluating $H_A$ holographically, we can substitute into \eq{modular H-ball} the holographic energy density $T_{tt}$ evaluated via the method of \cite{Balasubramanian:1999re,deHaro:2000vlm}.

    To extract some constraints from the relative entropy, we are interested in the case where $\sigma$ is the CFT vacuum state, and $\rho$ is a perturbative state (by either excitation or deformation)  away from $\sigma$, i.e.,
\be
\rho = \sigma + \delta \rho\;,  \qquad  |\delta \rho|/|\rho| <<1\;.
\ee 
Since the relative entropy takes its extremum at $\rho=\sigma$, its first order variation vanishes, which then yields the first law of entanglement thermodynamics \cite{Bhattacharya:2012mi,Faulkner:2013ica,Blanco:2013joa}, i.e.,
\be\label{1st law}
\Delta \langle H_A \rangle|_{\mathcal{O}(\delta \rho)} =\Delta S_A|_{\mathcal{O}(\delta \rho)}\;.
\ee
 On the other hand, the positivity of the second order variation will impose some unitarity bound on the deformed states of CFT.

    In the context of AdS/CFT correspondence, a dual state is characterized by an asymptotic AdS bulk metric, i.e., in the Poincare coordinates,
\be
ds^2={\ell^2 \over z^2} \Big(G(z) dz^2+ H_{\mu\nu}(z,x^{\mu}) dx^{\mu} dx^{\nu}\Big)
\ee
where $\ell$ is the AdS radius. The CFT vacuum state $\sigma$ corresponds to $G(z)=1$ and $H_{\mu\nu}=\eta_{\mu\nu}$. The perturbative state $\rho$ will be given by some slightly deviated $H_{\mu\nu}$ as well as $G(z)$. 

   The stress tensor $T_{\mu\nu}$ for each dual CFT state can then be evaluated holographically in the standard way as given in \cite{Balasubramanian:1999re, deHaro:2000vlm}. Therefore, we can then obtain the modular Hamiltonian $H_A$ on the region $A$ from $T_{\mu\nu}$ for the corresponding state  by the relation \eq{modular H-ball} if $A$ is a ball.  By evaluating $H_A$ for both $\rho$ and $\sigma$ states and subtracting them, we can then obtain $\Delta \langle H_A \rangle$.

   We still need to calculate $\Delta S_A$ holographically in order to obtain the relative entropy by \eq{relativeE-2}. The entanglement entropy can be obtained holographically by the Ryu-Takayanagi formula \cite{Ryu:2006bv,Ryu:2006ef}: 
\be \label{RT formula}
S_A = {\mbox{Area}(\gamma_A) \over 4 G_N}
\ee
where $\gamma_A$ is the co-dimensional two extremal surface ending on the entangling surface $\partial_A$ at the AdS boundary $z=0$. For static metric, $\gamma_A$ is simply the minimal surface on a fixed time slice.  We can then evaluate the $\gamma_A$ and thus $S_A$ for the bulk metrics corresponding to states $\rho$ and $\sigma$, and then subtract them to obtain $\Delta S_A$.

\subsection{Summary of Our Results}
 In this work we consider the  deformed holographic CFT with fermionic sources in AdS$_4$ space in the framework of Einstein-Cartan gravity with a postulated coupling constant for the bilinear interaction between axial current and torsion. Note that this coupling cannot be fixed by the dynamical symmetry of holographic CFT. Instead we find that the positivity of relative entropy disfavors such a coupling. 
 
 By turning on both normalizable and non-normalizable fermionic zero modes, the bulk geometry will be backreacted by the torsion at the second order of gravitational coupling.  We will solve the backreacted metric in the expansion of Newton constant, denoted by $\sk$, up to $\sk^2$ order, i.e., 
\be\label{metric2nd}
{\bf g}={\bf g}_0 + \sk \; {\bf g}_1 + \sk^2\; {\bf g}_2 
\ee 
where ${\bf g}_0$ is the pure AdS$_4$ metric. Note that ${\bf g}_1$ and ${\bf g}_2$ will encode the information about the fermionic sources.

Moreover, a bulk fermion of mass $m$ considered here is dual to a CFT operator of conformal dimension $\Delta$ given by \cite{Iqbal:2009fd}
\be\label{fermionmass}
\Delta={d\over 2}+ |m|\;. 
\ee
Unlike the case for the scalar field/operator, from \eq{fermionmass} we see no analogue of BF bound \cite{Breitenlohner:1982jf} on $m$ for preventing from instability.

In this paper, we will consider the backreaction due to the non-vanishing stress tensor of the bulk fermionic zero modes.  This is dual to the holographic deformed CFT given by \eq{couplingterm} and \eq{Ovev}. Then, we will evaluate the relative entropy up to $\sk^2$ (or $\delta \l^2$) order by using the backreacted metric.

 As we will see, the torsion will affect ${\bf g}_2$ but not ${\bf g}_1$. Thus, at the $\sk$ order, everything should go as the usual Einstein gravity so that the relative entropy evaluated holographically will satisfy the first law \eq{1st law}, i.e.,
\be\label{1st-order-rel}
S(\rho||\sigma)|_{\sk}=\Delta \langle H_A \rangle|_{\sk}-\Delta S_A|_{\sk}=0\;,
\ee
as expected. 
 
  For the evaluation of the second order relative entropy, we first need to evaluate $\Delta \langle H_A \rangle|_{\sk^2}$ by plugging $\langle T_{tt} \rangle|_{\sk^2}$ into \eq{modular H-ball}. However, following the method of \cite{Balasubramanian:1999re,deHaro:2000vlm}  the holographic evaluation of $\langle T_{tt} \rangle|_{\sk^2}$ via  ${\bf g}_2$  yields zero. Thus, we have  $\Delta \langle H_A \rangle|_{\sk^2}=0$.  Combined this result with the one of \eq{1st-order-rel}, we obtain  the relative entropy  \eq{relativeE-2} up to $\sk^2$ order, that is
\be  \label{relativeE-master}
 S(\rho||\sigma)=-\Delta S_A|_{\sk^2}\ge 0\;.
\ee

To evaluate $\Delta S_A|_{\sk^2}$, we need to first evaluate the the minimal surface $\gamma_A$ with respect to the metric ${\bf g}_0+ \sk {\bf g}_1$ up to $\sk^2$ order,  and we denote it by $\gamma_A^{(1)}$.  Then, we can obtain  $\Delta S_A|_{\sk^2}$ by
\be
\Delta S_A|_{\sk^2}:={\mbox{Area}(\gamma_A^{(1)})|_{\bf g} \over 4 G_N}|_{\sk^2}
\ee 
where $\mbox{Area}(\gamma_A^{(1)})|_{\bf g}$ means evaluating the area of surface $\gamma_A^{(1)}$ with respect to the metric \eq{metric2nd}.

In this work we find that the constraint \eq{relativeE-master} yields a constraint on $m$:
\be\label{m2l2}
m^2\ell^2 \ge {2\eta_t^2 \over \mu_0^2}
\ee
where $\eta_t$ is the postulated coupling constant for the bilinear interaction between bulk axial current and torsion, and $\mu_0$ characterizes the equation of state for the dual deformed fermion state  by
\be\label{EOState-intro}
P= {\mu_0 -2 \over 2 \mu_0}\; \varepsilon \;.
\ee
Here $P$ is the pressure and $\varepsilon$ is the energy density of the deformed fermion state.  Note that  the monotonicity condition ${\partial S(\rho||\sigma) \over \partial R_A}\ge 0$ yields the same condition \eq{m2l2}.

We see that the positivity of the relative entropy imposes constraint on the bilinear coupling $\eta_t$ in \eq{LpsiK} as well as the fermion mass $m$ and equation of state $\mu_0$. If there were no such coupling, i.e, $\eta_t=0$, then the relative entropy is positive for all $m$, which is consistent with the fact there is no BF-bound for fermion's mass and seems more natural. Otherwise, the constraint is on all three parameters in a nontrivial way.  It is interesting to see how this constraint can be understood as some energy condition in the bulk Einstein-Cartan gravity.

\section{Einstein-Cartan Gravity: A Brief Review}\label{sec 2}

    We will consider the backreaction to the bulk metric by turning on the bulk fermion matters. The nontrivial fermion matters will in general source torsion so that the gravity is naturally described by the  Einstein-Cartan theory. Here we will follow the formulation of four-dimensional torsional gravity with fermion matters given in \cite{Hehl:1976kj} (or \cite{Shapiro:2001rz}) to derive the field equations.

In the Riemann-Cartan spacetime, the torsion tensor is defined as 
\be
S_{\m \n}^{~~\,\l} = \frac{1}{2} ( \G_{\m \n}^\l - \G_{\n \m}^\l ) := \G_{[\m\n]}^\l \,,
\ee
in which $\G_{\m \n}^\l$ is the affine connection compatible to the metric
\be
\nb_{\s} \,g_{\m \n} = 0\,, 
\ee
where $\nb_{\s}$ the covariant derivative defined in terms of $\G_{\m \n}^\l$. This affine connection is specified uniquely in terms of metric and torsion:
\be
\G_{\m \n}^\l = {\tilde \G}_{\m \n}^\l - K_{\m \n}^{~~\,\l}\,,
\ee
where ${\tilde \G}_{\m \n}^\l$ is the Christoffel symbol computed from the metric, and $K_{\m \n}^{~~\,\l}$ is the so called contorsion tensor
\be
K_{\m \n}^{~~\,\l} = - S_{\m \n}^{~~\,\l} + S_{\n ~\, \m}^{~\,\l} - S_{~\,\m \n}^{\l} \,. 
\ee

  A general action for the four-dimensional torsional gravity with matter fields takes the following form:
\be
I = \int d^4 x \left[ \cL_M(\ps, \p\ps, g, \p g , S) + \frac{1}{2\k^2} \cL_R(g, \p g, S, \p S) \right]\,, 
\ee
where $\k^2 := 8 \pi G_N$. By varying the action, the formal field equation for matter is 
\be
\frac{\d \cL_M}{\d \ps} = 0 \,, \label{ome}
\ee
and the ones for the metric and torsion are
\be
- {1\over \ve} \,\frac{\d \cL_R}{\d g_{\m \n}} = \k^2  \,\s^{\m \n} \,,   \qquad
- {1\over \ve} \,\frac{\d \cL_R}{\d S_{\m \n}^{~~\,\l}} = 2 \eta_t \k^2  \,\m_{\l}^{~\, \n \m} \,, \label{ofe2}
\ee
where $\d \cL(Q, \p Q) / \d Q := \p \cL / \p Q - \p_\m \left[ \p \cL / \p (\p_\m Q) \right] $\,. We have introduced the metric energy-momentum tensor 
\be\label{sigmadef}
\s^{\m \n} := {2 \over \ve} \,{\d \cL_M \over  \d g_{\m \n}}
\ee
as well as a tensor with the meaning of a spin energy potential 
\be\label{mutensor}
\eta_t  \, \m_{\l}^{~\, \n \m} := {1 \over \ve}\, {\d \cL_M \over \d S_{\m \n}^{~~\,\l}}\,. 
\ee

In the above, we have introduced the coupling constant $\eta_t$ between the fermion and the torsion.  For $\eta_t=1$ we call such coupling canonical, which means that the fermion couples to the torsion with the same strength as its coupling to the non-torsion part of the affine connection, much like what metric does.  

   Explicitly, in this paper we will consider the fermion Lagrangian with generic coupling constant $\eta_t$, i.e.,
\be\label{Faction-1}
\cL_M= \frac{1}{2} \ve \, \left[ (\nb_\m {\bar \ps}) \g^\m \ps - {\bar \ps} \g^\m \nb_\m \ps - 2 m {\bar \ps} \ps \right] 
\ee
with the covariant derivative acting on $\psi$  given by 
\be\label{covDsplit}
\nb_\m = \p_\m + \frac{1}{4} \tilde{\om}_{\m}^{~\, ab} \g_{ab} +  { \eta_t \over 4}\, K_{\m\n\l} \gamma^{\n}\gamma^{\l}
\ee
where the Riemannian part of spin connection is defined as 
\be
\tilde{\om}_{\m~\,b}^{~\, a} = e_{\n}^{~ \,a} \,e^{\l }_{~\,b} \,\tilde{\G}^\n_{\m \l} - e^{\l}_{~\, b} \,\p_\m e_{\l }^{~\,a}\,.
\ee   

 The convention of gamma matrices adopted here is $\{\g_a ,\g_b\} = 2 \et_{ab} $ with $\et_{ab}={\mbox {diag}}\{-1, \,1, \,1, \,1 \}$,  $\g_5 = \g_0 \g_1 \g_2 \g_3$ and ${\bar \ps} = \ps^{\dag} (i \g_0) $. Latin indices label the components in the tangent space and Greek indices label that of the curved space. Note that $\g_\m = e_{\m}^{~\, a} \g_a$.  

  We can further separate the Riemannian part of  \eq{Faction-1} from the torsion part, i.e.,
\be\label{Faction-2}
\cL_M = { \tilde \cL}_M  + \eta_t \ve \, \t^{\m \n \l} \,K_{\l \n \m}\,, 
\ee
where ${ \tilde \cL}_M$ is the usual fermion Lagrangian with minimal coupling to the Riemannian part of spin connection.  Here we have introduced the spin angular momentum tensor $\t_{\l}^{~\, \n \m} $ by 
\be\label{spintaudef}
\eta_t  \,\t_{\l}^{~\, \n \m} :={1\over \ve} {\d \cL_M \over  \d K_{\m \n}^{~~\,\l}}
\ee
which is related to $\m_{\l}^{~\, \n \m}$ defined in \eq{mutensor} by $\t^{\m \n \l} =\m^{[\n \m] \l} $. 
From \eq{Faction-1} \eq{covDsplit} \eq{spintaudef} we obtain 
\be
\t^{\m \n \l } = \,\frac{1}{4} \,{\bar \ps} \g^{[\m} \g^{\n} \g^{\l]} \ps \,. \label{tdrc}
\ee
 It is easy to see that $\t^{\m \n \l }$ is directly related to the axial current $j^{\mu}_5:=\bar{\psi}\gamma^{\mu} \gamma_5 \psi$. Thus the second term on the R.H.S. of \eq{Faction-2} is the bilinear coupling between axial current and torsion given in \eq{LpsiK}.

    On the other hand, for the metric we will consider the Einstein-Cartan gravity, i.e., 
\be
\cL_R = \ve \, (R - 2 \Ld)
\ee
with $R = g^{\m \n} R_{\m \n}$, $R_{\m \n} = R_{ \l \m \n}^{~~~~\l}$ and 
\be
R_{ \m \n \l }^{~~~~\t}= 2 \,\p_{[ \m} \G^{\t}_{\n ] \l} + 2\, \G^{\t}_{[\m | \et} \,\G^{\et}_{ | \n ] \l} \,. 
\ee
Note that the Ricci tensor $R_{\m \n}$ is asymmetric in general, so is the Einstein tensor $G_{\m \n} := R_{\m \n} - 1/2 \,g_{\m \n} R$\,.  Also, $\Lambda$ is the cosmological constant and in the following we will take $\Lambda=-3/\ell^2$ where $\ell$ is the curvature radius of AdS space. 

A rather lengthy calculation gives
\bea
\frac{1}{\ve} \,\frac{\d \cL_R}{\d g_{\m \n}} &=&
- \,G^{\m \n} - \Ld g^{\m \n}+ {\hat \nb}_\l \left( T^{\m \n \l} - T^{\n \l \m} + T^{\l \m \n}  \right) \,,  \\
\frac{1}{\ve} \,\frac{\d \cL_R}{\d S_{\m \n}^{~~\,\l}}  &=&
- \,2  \left(\, T_\l^{~\m \n} -  T^{\m \n}_{~~~\l} + T^{\n ~ \m}_{~\,\l} \, \right) \,, 
\eea
where the modified divergence ${\hat \nb}_\l := \nb_\l + 2\, S_{\l \t}^{~~\,\t}$ and the modified torsion tensor  $T_{\m \n}^{~~\,\l} := S_{\m \n}^{~~\,\l} + 2 \,\d^\l_{[\m}\, S_{\n ] \t}^{~~~\t} $ have been introduced.

Substitute the above results  into (\ref{ofe2}), we get the field equations 
\bea
G^{\m \n} + \Ld g^{\m \n}&=& \k^2 \, \S^{\m \n}\,,  \label{fe1} \\
T^{\m \n \l} &=& \eta_t \k^2 \, \t^{\m \n \l}\,,   \label{fe2}
\eea
where we have introduced the asymmetric total energy momentum tensor 
\be\label{asymSigmadef}
\S^{\m \n} =\s^{\m \n} - \eta_t {\hat \nb}_\l \, \m^{\m \n \l}\;.
\ee
From \eq{sigmadef}, \eq{mutensor} and \eq{Faction-2} we have
\be
\S_{\m \n} = -\frac{1}{2} \left[ (\nb_\m {\bar \ps}) \g_\n \ps - {\bar \ps} \g_\n \nb_\m \ps  \right] \,. \label{Sdrc} 
\ee

   In fact, we can split the Einstein tensor  $G^{\m \n}$ in \eq{fe1} into its Riemannian part ${\tilde G}^{\m \n}$ as well as its non-Riemannian part and substitute the torsion terms in the latter part by means of  (\ref{fe2}), which is an algebraic relation. Then, we can obtain the combined field equation 
\be
{\tilde G}^{\m \n} + \Ld g^{\m \n} = \k^2 \, {\tilde {\bm \s}}^{\m \n}   \label{cfe}
\ee
with the combined energy momentum tensor
\begin{multline}
{\tilde {\bm \s}}^{\m \n} = {\bm \s}^{\m \n}  + \, \eta_t^2 \k^2  \, 
\left[ -4\, {\bm \t}^{\m \l}_{~~~[\t} \, {\bm \t}^{\n \t}_{~~~ \l]} -2 \, {\bm \t}^{\m \l \t} \,{\bm \t}^{\n}_{~\,\l \t} + {\bm \t}^{\l \t \m} \,{\bm \t}_{\l \t}^{~~\,\n}  \right. \\
 \left.  +  \frac{1}{2} \, g^{\m \n} \left( 4 \, {\bm \t}^{~\l}_{\et \,~[\t } \,{\bm \t}^{\et \t}_{~~~\l]}  + {\bm \t}^{\et \l \t} \,{\bm \t}_{\et \l \t}  \right) \right] \,,
   \label{cbt}
\end{multline}
which is symmetric by definition and obeys the conservation law ${\tilde \nb}_{\n}\, {\tilde {\bm \s}}^{\m \n} = 0$ (where ${\tilde \nb}$ is defined in terms of Christoffel symbols). For the fermion Largarangian \eq{Faction-2}  we have
\be
{\tilde {\bm \s}}_{\m \n} = {\tilde {\bm \S}}_{(\m \n)} - \frac{1}{2} \eta_t^2 \k^2 \,g_{\m \n} \, {\bm \t}^{\a \b \g} \,{\bm \t}_{\a \b \g} \,. \label{cbtdrc}
\ee
where ${\tilde {\bm \S}}_{\m \n}$ is the Riemannian part of $ {\bm \S}_{\m \n}$. 

The Dirac equation $\d \cL_M / \d {\bar \ps} =0$ takes the following form after using the second field equation (\ref{fe2}) :
\be
\g^\m {\tilde \nb}_\m \ps + \frac{3}{8} \eta_t \k^2 ({\bar \ps} \g_5 \g^{\m} \ps) \g_5 \g_\m \ps  + m \ps =0 \,.  \label{drc}
\ee

   In summary, the field equations we are going to solve are \eq{cfe} and \eq{drc} in the familiar Riemannian form but with the sources due to the fermion matter and the torsion given by (\ref{tdrc})  (\ref{Sdrc}) and (\ref{cbtdrc}).
      
As a matter of fact, the $\S^{\m \n}$ and $\t^{\m \n \l}$ defined by \eq{asymSigmadef} and \eq{tdrc} are the canonical energy-momentum tensor and spin angular momentum tensor. That is, if the matter equation (\ref{ome}) is fulfilled, the Noether's theorem indicates the following identifications
\be\label{stress&spin}
\ve \, \S_{\m}^{~\,\n} = \cL_M \,\d^{\n}_{\m} - \frac{\p \cL_M}{\p (\p_\n \ps)} \nb_\m \ps \,,  \qquad
\ve \, \t^{\m \n \l} = \frac{\p \cL_M}{\p (\p_\l \ps)} h^{[\n \m]} \ps \,,
\ee
and the conservation laws 
\be
{\hat \nb}_\n \,\S_{\m}^{~\,\n} = 2\, \S_{\l}^{~\,\n} S_{\m \n}^{~~\, \l}  + \t_{\n \l}^{~~\,\t} R_{\m \t}^{~~\,\n\l}\,, \qquad
{\hat \nb}_\l \,\t_{\m \n}^{~~\, \l}  = \S_{[\m \n]} \,, 
\ee
where $h^{\m \n}$ are the representation matrices of the infinitesimal coordinate transformation appropriate to the matter field.

\section{Solving Field Equations} \label{sec 3}

\subsection{Perturbative Field Equations}

   As the full solutions of the field equations in the presence of fermion matters and torsion are difficult to obtain even numerically, in this subsection we will first iteratively expand the field equations up to the second order of gravitational coupling constant. Then, we will solve them in the next subsection. 
    
    Naively we would like to solve the field equations order by order in the perturbative series of $\k^2$. However, as $\kappa^2$ is dimensionful, we will instead introduce an IR length scale $r_L$ and later perform the expansion with respect to a very small dimensionless parameter $\sk$ in the low energy limit, with 
\be
\sk:={\k^2 \over r_L^2}\;.
\ee
Note that $r_L$ is not an emerging IR scale but an intrinsic IR cutoff, which can be naturally thought as the overall size cutoff for the perturbative zero mode solutions. Moreover, as we will see that the resultant physical quantities such as entanglement entropy, modular Hamiltonian and also the derived information inequality will not depend on $r_L$. Thus, it is indeed not an emerging scale. 

To solve the field equations  \eq{cfe} and \eq{drc} perturbatively, we expand all the tensor fields in the powers of $\sk$ in the following manner:
\be
{\mathbf A} = {\mathbf A^{(0)}}  + \sk \; {\mathbf A^{(1)}} + \sk^2 \; {\mathbf A^{(2)}} + \cdots\;,
\ee
where ${\mathbf A}$ can be $g_{\mu\nu}$, $e^a_{\mu}$, ${\tilde \om}_{\m}^{~\, ab}$, ${\tilde \nb}_\m$, $\psi$, $\tilde{G}_{\mu\nu}$, $\tilde{\sigma}_{\mu\nu}$, $\cdots$. 

For our purpose we will solve up to $\sk^2$ order. The zeroth order field equations  are simply 
\bea
{\tilde G}^{(0)}_{\m \n} + \Ld g^{(0)}_{\m \n} &=& 0 \,,  \\
\left( \g^{\m(0)} {\tilde \nb}_\m^{(0)}   + m \right)\ps^{(0)} &=& 0 \,.
\eea
Note that there are vielbeins hidden in $\g^\m$'s, and $ {\tilde \nb}_\m^{(0)} $ is defined in terms of the 
zeroth order vielbeins: 
\be
{\tilde \nb}_\m^{(0)} = \p_\m + \frac{1}{4} {\tilde \om}_{\m}^{~\, ab\, (0)} \g_{ab} \,.
\ee

\vspace{4mm} 

The first order field equations  are
\bea
{\tilde G}^{(1)}_{\m \n} + \Ld g^{(1)}_{\m \n} &=&  r_L^2\, {\tilde {\bm \s}}^{(0)}_{\m \n} \,,  \label{einstein1} \\  ~ && \nn \\
\left( \g^{\m(1)} {\tilde \nb}_\m^{(0)}   +  \g^{\m(0)} {\tilde W}_\m^{(1)} \right)\ps^{(0)}  
+ \left( \g^{\m(0)} {\tilde \nb}_\m^{(0)}   + m \right)\ps^{(1)}    &&  \nn \\
+ ~\frac{3}{8}\, \eta_t^2 r_L^2 \, ({\bar \ps}^{(0)}  \g_5 \g^{\m (0)} \ps^{(0)} ) \g_5 \g_\m^{(0)}  \ps^{(0)}  &=& 0 \,, \label{dirac1}
\eea
in which
\bea
 {\tilde W}_\m^{(1)}  &:=&  \frac{1}{4} {\tilde \om}_{\m}^{~\, ab \,(1)} \g_{ab}  \,, \\ ~ && \nn \\
  {\tilde \s}^{(0)}_{\m \n}  &=&  {\tilde \S}_{(\m \n)} ^{(0)} =
 -\frac{1}{2} \left[ ({\tilde \nb}_{(\m}^{(0)} {\bar \ps^{(0)}}) \g_{\n)}^{(0)} \ps^{(0)}
  - {\bar \ps}^{(0)} \g_{(\n}^{(0)} {\tilde \nb}_{\m)}^{(0)} \ps^{(0)}  \right]  \,. \label{sigma0}
\eea
In the above, the notation $ {\tilde \nb}_{\m}^{(0)} { \bar \ps^{(0)}}$ is understood as $ \left({\tilde \nb}_{\m}^{(0)} {\ps^{(0)}}\right)^{\dag} (i\g_0)$.  

Obviously, from (\ref{einstein1}) we see that the first order perturbed metric  $g^{(1)}_{\m \n}$ is sourced  only by  the stress tensor of the zeroth order fermion field. To see the influence of the torsion on the metric, we need to look into the  second order perturbation $g^{(2)}_{\m \n}$ which satisfies  
\be\label{2ndEinstein}
{\tilde G}^{(2)}_{\m \n} + \Ld g^{(2)}_{\m \n} =  \,  r_L^2 \, {\tilde \s}^{(1)}_{\m \n} \,,
\ee
where 
\bea \label{sigma1}
{\tilde \s}^{(1)}_{\m \n} &=&  {\tilde \S}_{(\m \n)} ^{(1)}  
- \frac{1}{2} \, \eta_t^2  r_L^2\, g_{\m \n}^{(0)} \, \t^{a b c (0)} \,\t_{a b c}^{(0)}  \,,  \\ ~ && \nn \\
\t^{a b c (0)}  &=& \,\frac{1}{4} \,{\bar \ps}^{(0)}  \g^{[a} \g^{b} \g^{c ]} \ps^{(0)} \,, \label{tau0}\\ ~&&  \nn  \\
 {\tilde {\bm \S}}_{(\m \n)} ^{(1)}    &=&  
 -\frac{1}{2} \left[ ({\tilde \nb}_{(\m}^{(0)} {\bar \ps^{(0)}}) \g_{\n)}^{(0)} \ps^{(1)}
  - {\bar \ps}^{(0)} \g_{(\n}^{(0)} {\tilde \nb}_{\m)}^{(0)} \ps^{(1)}  \right]   \nn \\
  &&   -\frac{1}{2} \left[ ({\tilde \nb}_{(\m}^{(0)} {\bar \ps^{(1)}}) \g_{\n)}^{(0)} \ps^{(0)}
  - {\bar \ps}^{(1)} \g_{(\n}^{(0)} {\tilde \nb}_{\m)}^{(0)} \ps^{(0)}  \right]   \nn  \\
  &&  -\frac{1}{2} \left[ ({\tilde \nb}_{(\m}^{(0)} {\bar \ps^{(0)}}) \g_{\n)}^{(1)} \ps^{(0)}
  - {\bar \ps}^{(0)} \g_{(\n}^{(1)} {\tilde \nb}_{\m)}^{(0)} \ps^{(0)}  \right]    \label{bigsigma1} \\
  &&  -\frac{1}{2} \left[ ({\tilde W}_{(\m}^{(1)} {\bar \ps^{(0)}}) \g_{\n)}^{(0)} \ps^{(0)}
  - {\bar \ps}^{(0)} \g_{(\n}^{(0)} {\tilde W}_{\m)}^{(1)} \ps^{(0)}  \right]  \,.  \nn
\eea
Similarly, the notation ${\tilde W}_{\m}^{(1)} {\bar \ps^{(0)}}$ is understood as $\left({\tilde W}_{\m}^{(1)} {\ps^{(0)}}\right)^{\dag} (i\g_0)$.  From \eq{sigma1} and \eq{2ndEinstein} we see that the spin angular momentum tensor $\t^{a b c (0)}$ which is related to the modified torsion tensor $T^{\m\n\l}$ \eq{fe2} now comes into play as part of the source for the perturbation $g^{(2)}_{\m \n}$.

\vspace{3mm}

\subsection{Solutions} \label{sec 4}

   We will now solve the above perturbative field equations up to $\sk^2$ order. Then, we can use these solutions to evaluate the correction to the holographic entanglement entropy via Ryu-Takayanagi formula.

\vspace{2mm}

\subsubsection{The Zeroth Order Fermion Solution} 

\vspace{2mm}

For simplicity we drop the superscript of $\psi^{(0)}$ throughout this sub-subsection. The zeroth order Dirac equation is 
\be
\g^\m \left( \p_\m + \frac{1}{4} {\tilde \om}_{\m}^{~\, ab} \g_{ab}  \right) \ps  + m \ps = 0\,.
\ee
We choose the 0$^{th}$-order metric to be the pure AdS$_4$: 
\be
ds^2 = \frac{r^2}{\ell^2} \left( -dt^2 + dx^2 + dy^2 \right) + \frac{\ell^2}{r^2} dr^2 \,.
\ee
 
Writing 
\be
\ps(t,r,x^i) = e^{-i \om t \,+\, i k_1 x \,+\, i k_2 y} \ph(r)\,,
\ee
the Dirac equation is turned into
\be \label{diracmom}
\frac{r}{\ell} \,\g_3 \,\p_r \ph  + i\, \frac{\ell}{r}\, \g_i \, k^i \, \ph + \frac{3}{2\ell} \g_3 \,\ph + m\ph  = 0\,,
\ee
where $\{k^i\} = \{\om, k_1, k_2\}$ and $\g_i = \{\g_0, \g_1, \g_2\}$.  
We work in the following representation of the Dirac matrices:
\be \label{chga}
 \g_{i} = \left(\begin{tabular}{cc}0 & ${\tilde \g}_{i}$ \\ ${\tilde \g}_{i}$ & 0\end{tabular}\right), \qquad  
 \g_3 = \left(\begin{tabular}{cc}$1$ & $0$ \\ $0$ & $-1$\end{tabular}\right) 
\ee
where ${\tilde \g}_{i}$ are the 2$\times$2 gamma matrices for the (2+1)-dimensional dual boundary theory: 
${\tilde \g}_{0} = i \s_2$,  ${\tilde \g}_{1} = \s_1$, ${\tilde \g}_{2} = \s_3$ with $\s_i$'s the Pauli matrices.  

Decompose $\ph$ into
\be
\ph = \left( \begin{array}{c} \ph_+ \\ \ph_- \end{array}\right)\;,
\ee 
the equation (\ref{diracmom}) can be reduced to a set of coupled equations 
\bea
\left( \frac{r}{\ell} \, \p_r + \frac{3}{2\ell} + m \right) \phi_+ \,+\, i \,\frac{\ell}{r} \,{\tilde \g} \cdot k \,\phi_- &=& 0 \label{cp1}\;, \\
\left( \frac{r}{\ell} \, \p_r + \frac{3}{2\ell} - m \right) \phi_- \,-\, i\,\frac{\ell}{r} \,{\tilde \g} \cdot k \,\phi_+ &=& 0 \label{cp2}\;.
\eea
From the above we can obtain a second order equation of $\phi_+$ which is the deformed Bessel equation, i.e.,
\be\label{besseleq}
r^2 \phi_+^{''} + \left[ 1 - 2\cdot(-2) \right] r \phi_+^{'}  + \left[ -k^2 \ell^4 r^{-2} + 4 - \left(m\ell-\frac{1}{2}\right)^2 \, \right] \phi_+ = 0\,.
\ee
After solving $\ph_+$, one can then obtain $\ph_-$ from \eq{cp1}. The explicit solutions are listed in the appendix. 

   In this paper, for our purpose we will consider the simpler case with $\om = k_1 = k_2 = 0$. Otherwise, the backreacted metric will be non-stationary and inhomogeneous along the traverse directions (see appendix A for a brief discussion on the case with non-vanishing $\om, \,k_1,\,k_2$). Thus, we will avoid this kind of complications when considering the holographic entanglement entropy in the backreacted background geometry.  For such a case, the equations (\ref{cp1}) and (\ref{cp2}) are decoupled, and the solution is simply
\be\label{0mode}
\psi^{(0)}  =  \left( \begin{array}{c} r^{-3/2 \,-\,  m \ell} \,a_+ \\  r^{-3/2 \,+\, m \ell} \,a_- \end{array}\right)\,,
\ee
where $a_+$ and $a_-$ are two independent arbitrary constant 2-component spinors.   

\vspace{2mm}


\subsubsection{The First Order Metric Solution}

\vspace{4mm}

  We now use the zeroth order fermion solution \eq{0mode} to source the first order perturbation of the metric. Without loss of generality we choose the constant spinors to be 
\be \label{IntC}
 a_+ = \{0,\, \a\}^{\mathrm{T}}, \qquad a_- = \{i \b,\, 0\}^{\mathrm{T}}
\ee  
where $\a$, $\b$ are real constant numbers. The corresponding  energy-momentum tensor ${\tilde {\bm \s}}_{\m \n}^{(0)}$ can be worked out through (\ref{sigma0}), and the result is 
\be\label{sigma0-1}
{\tilde {\bm \s}}_{\m \n}^{(0)} = {\mbox {diag}} \left\{ 0\,, ~ 0\,, ~ 0\,, ~ -\frac{2  m\ell^2 \a \b }{r^5} \right\}\,.
\ee
Note that integration constant chosen in \eq{IntC} is prescribed in \cite{Iqbal:2009fd} when considering the holographic dynamics of the fermionic operators with nontrivial transports.  This is also reflected in the fact that ${\tilde {\bm \s}}_{\m \n}^{(0)}$ is nonzero only if both $\alpha$ and $\beta$ are nonzero. Otherwise, there will be no backreaction to the bulk so that the relative entropy at all orders are zero. This is in contrast to the cases for the bulk scalar as considered in \cite{Faulkner:2014jva}.  Thus, in the following we will consider the case with nonzero $\alpha$ and $\beta$.

If we assume $m< 0$ \footnote{In the case of $m>0$ the roles of $\alpha$ and $\beta$ as source and vacuum expectation value (vev) just swap as the normalizable and non-normalizable modes swap, too.  Therefore, we can just stick to $m<0$ in the discussion of the main text.}, then $\phi_+$ associated with $\alpha$ is the non-normalizable mode and  $\phi_-$ associated with $\beta$ the normalizable mode.  By the AdS/CFT dictionary, e.g. see \cite{Iqbal:2009fd,Liu:2009dm}, this is dual to deforming the CFT by the term shown in \eq{couplingterm} and \eq{Ovev} with the following identification
\be\label{singletrace}
\delta \lambda \longleftrightarrow \{0,~ \alpha\} (i {\tilde{\g}}_0)\;,  \qquad  \qquad   
\langle \int d^d x \; \mathcal{O}_{\Delta}(x) \rangle \longleftrightarrow \{ i \beta, ~0\}^{\mathrm{T}} \;.
\ee
Thus, our relative entropy is to compare some UV state with the corresponding state under the RG flow driven by \eq{couplingterm} and \eq{Ovev}. Formally, we will assume $\alpha$ and $\beta$ to be very small, i.e., by restricting to the nearby region of UV fixed point.

The first order field equation \eq{einstein1} is solved by the following metric ansatz
\bea\label{ansatz1}
\sk \,g_{\mu\nu}^{(1)}dx^{\mu}dx^{\nu} &=& \sk \,{ r^2 \over \ell^2 } \left[- b_t  \left({ r \over r_L}\right)^{-q_t} dt^2 
+  b_x \left({ r \over r_L}\right)^{-q_x} dx^2 +  b_y \left({ r \over r_L}\right)^{-q_y} dy^2\right]  \nn \\
&& + ~ \sk \,b_r \left({ r \over r_L}\right)^{-q_r} {\ell^2 \over r^2}  \,dr^2 
\eea
with $q_t = q_x = q_y = q_r = 3$ and arbitrary dimensionless constants $b_\m \, 's$ satisfying 
\be\label{bsum1}
b_t + b_x + b_y + b_r = {2  m \ell^2 \a \b \over 3 \,r_L} \,. 
\ee

 Naturally we choose $b_x = b_y$ so that the transverse space is homogeneous and isotropic. As a matter of fact, we will need $b_x = b_y = 0$ so that the holographic entanglement entropy is free of IR divergence (refer to appendix B for detailed analysis). Hence the first order backreacted metric is 
\be\label{g1}
\sk \,g_{\mu\nu}^{(1)}dx^{\mu}dx^{\nu} = - \sk \, b_t  \left({ r \over r_L}\right)^{-3} { r^2 \over \ell^2 }\,dt^2 
 +  \sk \,b_r \left({ r \over r_L}\right)^{-3} {\ell^2 \over r^2}  \,dr^2
\ee 
where
\be\label{brbt}
b_r =  \mu_0 {  m \ell^2 \a \b \over r_L }\,, \quad\quad\quad b_t = \left( {2 \over 3} - \mu_0  \right){  m \ell^2 \a \b \over r_L } \, ,
\ee
with $\m_0$ some undetermined integration constant characterizing the metric configuration. It is similar to the black hole mass parameter of the vacuum solution in Einstein gravity.


\subsubsection{The First Order Fermion Solution}

 We again consider only the zero-mode (i.e., $k^i=0$) for the first order fermion perturbation, denoted by $\ps^{(1)}$, which can be decomposed as follows:
\be
\ps^{(1)} = \left( \begin{array}{c} \ps^{(1)} _+ (r) \\~\\ \ps^{(1)} _- (r) \end{array}\right)\,.
\ee 
Then, the first order field equation for $\ps^{(1)}$ also decompose into 
\bea
\left( \frac{r}{\ell} \p_r + \frac{3}{2\ell} + m \right) \ps^{(1)} _+  + \frac{\Delta_+}{\ell} \, r^{-9/2 \,-\, m \ell} a_+ &=& 0\,,  \\
-\left( \frac{r}{\ell} \p_r + \frac{3}{2\ell} - m \right) \ps^{(1)} _-  + \frac{\Delta_-}{\ell} \, r^{-9/2 \,+\, m \ell} a_-  &=& 0 \,, 
\eea
where  
\be
\Delta_{\pm} =   {1\over 4} \left(3 \eta_t^2 \,+\, 2 \mu_0 m^2 \ell^2 \pm (3 \mu_0 - 2) m \ell \right) \ell \a \b r_L^2\;.
\ee

  It is straightforward to solve the above equations and the results are 
\bea
\ps^{(1)} _+ &=& \Delta_+ \left( \frac{1}{3}\, r^{-9/2 \,-\, m \ell} + v_1\, r^{-3/2 \,-\, m \ell}  \right) a_+ \,,\\
\ps^{(1)} _- &=& \Delta_- \left( -\frac{1}{3} \, r^{-9/2 \,+\, m \ell} + v_2\, r^{-3/2 \,+\, m \ell} \right) a_-\,,
\eea
where $v_1$, $v_2$ are integration constants for the ``homogeneous solution". Note that the ``homogeneous solution" has the same power of $r$ as for the zeroth order solutions \eq{0mode}, thus we can simply set $v_1$ and $v_2$ to zero \footnote{From the point of view of solving the full solutions, we need only two integration constants, which should be fixed by the proper boundary conditions. When solving in the perturbative framework, it seems that we have two integration constants at each order. However, we should lump these integration constants together up to the order we solve, and then fixed the two lumped integration constants by boundary conditions. This is equivalent to just fixing the zeroth order integration constants e.g, $\alpha$ and $\beta$, but setting all the higher order one to zero. In the context of dual CFT, the lumped integration constants, i.e., $\alpha$ and $\beta$ are just the renormalized source and vev.}. Thus, the first order fermion solution is 
\be
\ps^{(1)} =  \frac{1}{3} \left( \begin{array}{c} \,~\Delta_+ \,r^{-9/2 \,-\, m \ell} \,a_+ \\
-\Delta_- \, r^{-9/2 \,+\, m \ell} \,a_-\end{array}\right)\,.
\ee

\vspace{5mm}


\subsubsection{The Second Order Metric Solution}

   The first order metric and fermion solutions can then source the second order metric perturbation. Based on the results in the previous subsections we can obtain the corresponding energy-momentum and spin angular momentum tensors via  (\ref{sigma1}), (\ref{tau0}) and (\ref{bigsigma1}), and the results are 
\bea
{\tilde {\bm \s}}^{(1)}_{\m \n}  &=&  \left( \begin{array}{cccc} 
~-\frac{3 \,\a^2 \b^2 \eta_t^2 r_L^2}{4 \,r^4 \ell^2}~ & \frac{(3\mu_0 - 2) m \a^2 \b^2 r_L^2}{4 \,r^4 \ell} & 0 &0\\  ~\\
\frac{(3\mu_0 - 2) m \a^2 \b^2 r_L^2}{4\, r^4 \ell} &  \frac{3 \,\a^2 \b^2 \eta_t^2 r_L^2}{4 \,r^4 \ell^2}&0 & 0  \\  ~\\
0 &0 & \frac{3 \,\a^2 \b^2 \eta_t^2  r_L^2}{4 \,r^4 \ell^2}& 0 \\~\\
0 &0  & 0 &  - \frac{(9\eta_t^2 \,+\, 4 (9\mu_0 -2) m^2 \ell^2) \ell^2 \a^2 \b^2 r_L^2}{12 \,r^8}
 \end{array}\right)   \,.
\eea 
The torsion contributes to ${\tilde {\bm \s}}^{(1)}_{\m \n}$ through both the spin angular momentum $ \bm{\t}^{a b c (0)}$ and the first order fermion field $\ps^{(1)}$, since the latter is solved from  (\ref{dirac1}) containing a four-fermion interaction term which is due to the torsion coupling.

It turns out that the second order field equation \eq{2ndEinstein} is solved uniquely by the following second order metric perturbation: 
\be
\sk^2 g_{\mu\nu}^{(2)}dx^{\mu}dx^{\nu} = - \sk^2 \,k_t \left({r \over r_L}\right)^{-p_t} \frac{r^2}{\ell^2}\,dt^2 
\,+\, \sk^2 \,k_{tx} \left({r \over r_L}\right)^{-p_{tx}}  \frac{r^2}{\ell^2} \,dt \,dx
\,+\,  \sk^2 \,k_r \left({r \over r_L}\right)^{-p_r} \frac{\ell^2}{r^2}\,dr^2\,
\ee
with $p_t = p_r = p_{tx} = 6$ and 
\bea
k_t ~&=&~  \frac{(2 - 3 \mu_0)\, m^2 \ell^4 \a^2 \b^2 }{18\, r_L^2} \,,  \\
k_{tx} ~&=&~  \frac{(2 - 3 \mu_0)\, m \ell^3 \a^2 \b^2}{18\, r_L^2} \,, \\
k_r ~&=&~ \frac{(\eta_t^2 + 4 \mu_0^2 m^2 \ell^2) \ell^2 \a^2 \b^2}{4 \, r_L^2} \,.
\eea

Later on it is more convenient to work in the planar coordinate $z:=\ell^2/r$ with also $z_L :=\ell^2/r_L$ . Combined the above, the metric up to $\sk^2$ order  takes the form
\be\label{backreactg}
ds^2 = \frac{\ell^2 }{z^2} \left( - F(z) \,dt^2 + H(z)\, dt \,dx +  dx^2 + dy^2 + G(z)\, dz^2 \right) \,, 
\ee
with 
\bea 
F(z) &=& 1 + \sk\, b_t\, {z^3 \over z_L^3} + \sk^2\, k_t\, {z^6 \over z_L^6} \,, \nn\\
G(z) &=& 1+ \sk\, b_r\, {z^3 \over z_L^3} + \sk^2\, k_r\, {z^6 \over z_L^6} \,, \label{gzz} \\
H(z) &=& \sk^2\, k_{tx} \, {z^6 \over z_L^6} \,. \nn
\eea
We could evaluate the holographic stress tensor \cite{Balasubramanian:1999re, deHaro:2000vlm} for the above  backreacted metric  and the result is 
\be \label{stresstensor}
T_{\m \n} \,=\, \sk\, { \ell^2 \over 16 \pi G_N z_L^3  }\,{\mbox {diag}} 
\left\{ \,2 b_r \,, \, - (2 b_r + 3 b_t) \,, \, - (2 b_r + 3 b_t)   \,\right\}\,,
\ee
with $\m, \n = t, \,x, \,y$. It is straightforward to check that the traceless condition for 3d CFT  leads to $b_t + b_r =0$ which is in conflict with \eq{brbt} (for any finite $\mu_0$), indicating the violence of conformal symmetry due to the presence of the fermion source, i.e, the deformation. 

   Moreover, using  \eq{brbt}  we see that the parameter $\mu_0$ characterizes the equation of state of the stress tensor for the deformed fermion state by
\be\label{EOState}
P= {\mu_0-2 \over 2 \mu_0}\; \varepsilon
\ee
where $\varepsilon=T_{tt}$ and $P=T_{xx}=T_{yy}$.  We see that this equation of state approaches the one for conformal fluid if $\mu_0 \rightarrow \infty$  though the stress tensor is strictly not traceless.


\vspace{2mm}

\section{Relative Entropy for the Deformed Fermion State} \label{sec 5}

  Now we would like to extract the change of the holographic entanglement entropy  via Ryu-Takayanagi formula  \eq{RT formula} for the deformed fermion state described by the backreacted geometry \eq{backreactg} and \eq{gzz}.

We choose the region $A$ to be a disk of radius $R_A$\,, i.e., 
\be
A:=\{(x,y)| x^2 + y^2 \le R_A \}\,.
\ee
As our backreacted geometry \eq{backreactg} \eq{gzz} is in general stationary at the $\sk^2$ order, in principle we ought to solve for the extremal surface which is a bit tedious. However, since we are only interested in the resultant holographic entanglement entropy up to $\sk^2$ order, a simpler prescription \cite{Bhattacharya:2013bna} could be adopted as the following: firstly we solve for the minimal surface $\gamma_A$ in the geometry \eq{g1} and \eq{brbt} which is up to $\sk^2$ order and static, then we evaluate the area of the same surface in the metric \eq{backreactg} and \eq{gzz} which is up to $\sk^2$ order. We do not need to solve the the shape of extremal surface in the second order perturbed metric, since $\gamma_A$ is an extremal surface with respect to the first order perturbed metric.  

The area functional for a surface described by $r(z)$ in constant time slice is given by 
\be\label{areasphere}
\mbox{Area} = 2 \pi \ell^2 \int_{\epsilon}^{z_*} {dz \over z^2} \,r(z) \sqrt{r'(z)^2 + G(z)} \,, 
\ee
where we have turned $(x, y) \to (r, \th)$\,,  $\epsilon$ is the UV cutoff and $z_*$ is the turning point where $r'(z)$ diverges. It is straightforward to vary the area functional to get the equation of motion for the minimal surface $r(z)$: 
\be\label{surfaceeom}
z\, G(z)^2 + {1 \over 2}\, r(z)\, r'(z) \left( z\, G'(z) + 4\, r'(z)^2 \right) + G(z) \left( z\, r'(z)^2 + r(z) (2 \,r'(z) - z\, r''(z)) \right) = 0\,.
\ee
To solve for the minimal surface for the first order perturbed metric, we set $G(z)$ to be $G(z)^{(1)} = 1 + \sk\, b_r z^3 / z_L^3$\,, equation \eq{surfaceeom} could be solved perturbatively, i.e., 
\be\label{1rz}
r(z)^{(1)} = r_0(z) + \sk\, r_1(z) + \sk^2\, r_2(z) + \cdot\cdot\cdot
\ee
for our purpose we need to solve for the minimal surface up to $\sk^2$ order (so that it is still a minimum for a shape deformation under the second order metric perturbation, which is at least of $\sk^2$ order): 
\bea
r_0(z) &=&  \sqrt{z_*^2 - z^2}\;,   \\~\nn\\
r_1(z) &=& \frac{b_r \left(2 z_*^5 - z^3 (z_*^2 + z^2)\right)}{8 z_L^3 \sqrt{z_*^2 - z^2 \;}}\;,  \\~\nn\\
r_2(z) &=& \frac{b_r^2 \,J(z)}{4480 \,z_L^6 \sqrt{z_*^2 - z^2} \,(z_* + z)}  \;,
\eea
where
\begin{multline}
J(z) ~=~  447\, z_*^9 - 417\, z_*^8 z - 572\, z_*^7 z^2 
 + 152\, z_*^6 z^3 - 64\, z_*^5 z^4 + 356\, z_*^4 z^5    \\
 - 31\, z_*^3 z^6 - 31\, z_*^2 z^7 + 80\, z_* z^8 + 80\, z^9 - 864 \,z_*^8 (z_* + z) \log{\left(2 \,z_* \over z_* + z\right)}\,. 
\end{multline}
$z_*$ is related to $R_A$ by the relation $ R_A = r(0)^{(1)}$\,, from which we could solve $z_*$ in terms of $R_A$\,:
\be
z_* ~=~ R_A \,-\,  \sk \,{b_r \over 4\, z_L^3} R_A^4 \,+\, \sk^2 \,{b_r^2\, (673 + 864 \log{2})\over 4480\,z_L^6}  R_A^7 \,+\,  \cdot\cdot\cdot\,.
\ee

The holographic entanglement entropy for stationary metric given by \eq{backreactg} and \eq{gzz} is evaluated up to $\sk^2$  order  by substituting \eq{1rz} into \eq{areasphere} while setting $G(z)$ to be $G(z)^{(2)} = 1 + \sk\, b_r z^3 / z_L^3 + \sk^2 \, k_r z^6 / z_L^6$\,, the result is 
\be
 S_A = S_{0} + S_{1}  + S_{2}  + \cdot\cdot\cdot\,,
\ee
with
\bea
S_{0}  &=&  { \pi\, \ell^2 R_A \over 2 \, G_N\, \e} -  \frac{\pi \,\ell^2}{2 \,G_N }\,, \\~\nn\\
S_{1}  &=&  \mu_0 m \a \b\, { \,\pi^2  R_A^3 \over 2\, \ell^2}\,,  \\~\nn\\
S_{2} &=& (2\eta_t^2 - \mu_0^2 m^2 \ell^2) \a^2 \b^2 \,{4 \pi^3 G_N R_A^6  \over 35 \, \ell^8}\;, 
\eea
in the above we have replaced $\sk$ with $\k^2 / r_L^2 \equiv 8 \pi G_N z_L^2 / \ell^4$\,. 

On the other hand, the modular Hamiltonian could be evaluated from \eq{modular H-ball} and  \eq{stresstensor}. Note that there is only $\mathcal{O}(\sk)$ contribution in holographic stress tensor, 
\be
T_{tt}  = \sk \, {b_r \,\ell^2 \over 8 \pi G_N z_L^3} = \frac{\mu_0 m \a \b}{ \ell^2}\;,
\ee
\be
\langle H_A \rangle \equiv H_1 = 2 \pi \int_{r<R_A} r dr d\theta \,\frac{R_A^2 - r^2}{2 R_A}\, T_{tt} =  \mu_0 m \a \b \,\frac{\pi^2  R_A^3}{2 \,\ell^2} \;. 
\ee
From the above, we find that the relative entropy $S(\rho||\sigma):=\Delta \langle H_A \rangle - \Delta S_A$ vanishes at the first order because 
\be
H_1 = S_1\;.
\ee
This is the first law of entanglement thermodynamics as expected. 

  At the second order, the stress tensor is zero so that the relative entropy is nothing but $S(\rho||\sigma)=-S_2$. Then the positivity of the relative entropy gives 
\be\label{finalB}
m^2 \ell^2 \ge {2\eta_t^2 \over \mu_0^2} \;. 
\ee
Moreover, note that the monotonicity of the relative entropy, i.e., ${\partial S(\rho||\sigma) \over \partial R_A}\ge0$ gives the same constraint. 

Since there is no BF-like bound on the fermion mass $m$, it seems that a natural way to satisfy \eq{finalB} without fine-tuning $\eta_t$, $m$ and $\mu_0$ is to set $\eta_t=0$, which might indicate that the torsional coupling \eq{LpsiK} is disfavored hence the torsional gravity locates at the swampland \cite{Vafa:2005ui,Ooguri:2006in}. However, we will take the alternative interpretation that the positivity of relative entropy requires that such a bilinear interaction between axial fermion current and torsion could exist only for massive enough fermions. 

One might note that the change of the entanglement entropy $\Delta S_A \approx S_1 > 0$, which is in conflict with the intuitive expectation that monotonic decrease of entanglement entropy under RG flow \cite{Myers:2010tj,Jafferis:2011zi,Ishihara:2012jg,Rosenhaus:2014woa,Rosenhaus:2014nha,Ben-Ami:2015zsa}.   This is however not in conflict with the fact demonstrated in \cite{Swingle:2013zla, Cremonini:2013ipa} that the entanglement entropy does not monotonically decrease in the generic Lorentz-violating field theories. Though our deformation term \eq{couplingterm} is Lorentz invariant, the coupling source $\delta \lambda$ (or the vev of the deformation operator $\langle \int d^d x \; \mathcal{O}_{\Delta}(x) \rangle$) is a spinor so that its nonzero value violates the Lorentz invariance spontaneously. We thus expect $\Delta S_A \approx S_1 > 0$ (note that $S_2$ is one order smaller than $S_1$ and thus neglected.).

  On the other hand, the positivity and monotonicity of relative entropy, as they should be by the mathematical construction, have nothing to with Lorentz non-invariance. If the positivity condition is violated, it indicates that the underlying theory is pathological, i.e., swampland. In particular, implied by the first law of entanglement thermodynamics, the first oder relative entropy is zero even though  $S_1$ is not.   This is why one needs to go to second order for checking the positivity and monotonicity of relative entropy.

\vspace{2mm}

\section{Conclusion and Discussion}\label{sec 6}
    
        In this paper we study the positivity constraint of the relative entropy in the torsion gravity, which is dual to a deformed CFT by turning on a fermionic operator of nonzero vacuum expectation value. To consider the backreaction caused by the bulk fermion, one is free to postulate a conventional bilinear coupling between axial fermion current and torsion, which cannot be specified by the dynamical symmetry of CFT. Interestingly we find that the positivity condition imposes some constraint involving this coupling, fermion mass and equation of state.  Our results provide an example of exploiting the quantum information inequalities to constrain the holographic CFT with a novel torsion operator.
        
        We would like to emphasize that the relative entropy considered here for the deformed holographic CFT is different from the discussions in \cite{Lin:2014hva,Lashkari:2014kda,Lashkari:2015hha,Jafferis:2015del,Lashkari:2016idm} where the relative entropy is obtained by comparing the quantum states in the same CFT. 
In this general scope, we can have more handles to further explore the 
constraints in the dual gravity by the quantum information inequalities. 
          
        As discussed in \cite{Lashkari:2014kda,Lashkari:2015hha,Lashkari:2016idm}, the positivity of the relative entropy for the excited holographic state can be formulated as the positive energy condition in the bulk gravity. We believe that our positivity constraint of relative entropy should also correspond to some quasi-local energy condition in the Einstein-Cartan gravity. However, it has been known that the quasi-local energy in the torsion gravity is more subtle than in the Einstein gravity, and a commonly consented energy condition seems still lacking \cite{Nester}.  We will explore this problem within the framework of \cite{Iyer:1994ys,Hollands:2012sf} in a future work.

\vspace{2mm}
    
     \section*{Acknowledgement}
We would like to thank Bin Chen, Song He, Jia-Rui Sun, Ling-Yan Hung, James Nester and Yu Tian for very helpful discussions. This work is supported by Taiwan Ministry of Science and Technology through Grant No.~103-2112-M-003 -001 -MY3  and 103-2811-M-003 -024. BN is supported in part by NSFC under Grant No.~11505119.

       \vspace{2mm}
       
    
\section*{A.  Discussion on General Solutions with Nonzero $\omega$ and $\vec{k}$}  

For general non-vanishing parameters $\om, \,k_1, \,k_2$, the solution to equation \eq{besseleq} could be expressed in terms of Bessel functions  \cite{Iqbal:2009fd,Liu:2009dm} 

\be
\phi_+ (r) = r^{-2} Z_{m\ell - \frac{1}{2}} \left(\frac{\ell^2|k|} {r}\right) a_+ \,,
\ee
where $Z_\nu (z)$ is some Bessel function and $k^2 = -\om^2 + k_1^2 + k_2^2\,$\,, while $a _+$ is an arbitrary constant spinor with 2 components. The exact solution satisfying the in-falling boundary condition in IR regime is as the following:  
\be \label{phiplussolution}
\phi_+(r) =  \left\{ \begin{array}{cc} r^{-2} K_{m\ell - 1/2} \left( \frac{\ell^2\sqrt{k_1^2 + k_2^2 - \om^2}}{r} \right) a_+  \,,
& (\,k^2 > 0\,)  \\  
r^{-2} H^{(1)}_{m\ell - 1/2} \left( \frac{\ell^2\sqrt{\om^2 - k_1^2 - k_2^2 }}{r} \right) a_+  \,,
& ~~ (\,\om > \sqrt{k_1^2 + k_2^2}\,)  \\  
r^{-2} H^{(2)}_{m\ell - 1/2} \left( \frac{\ell^2\sqrt{\om^2 - k_1^2 - k_2^2 }}{r} \right) a_+  \,.
& ~~~~~ (\,\om < - \sqrt{k_1^2 + k_2^2}\,)
 \end{array}\right.
\ee 
We would be interested in the case with $\om > \sqrt{k_1^2 + k_2^2}$\,. 

\vspace{4mm}

The solution $\phi_-(r)$ should be related to $\phi_+(r)$ through (\ref{cp1})\,. Employing the following recursive relations of all kinds of Bessel functions: 
\bea
Z_{\nu -1} + Z_{\nu +1} &=& \frac{2\nu}{z} Z_\nu \,, \\
Z_{\nu -1} - Z_{\nu +1} &=& 2 \,Z'_\nu\,,
\eea
we obtain that 
\be
\phi_-(r) = r^{-2} Z_{m\ell + \frac{1}{2}} \left( \frac{ \ell^2 |k|}{r} \right) \left( \frac{i\, {\tilde \g} \cdot k}{|k|} \right) a_+\,.
\ee
For simplicity we will take the following form of shorthand: 
\be
Z_+ \equiv Z_{m\ell - \frac{1}{2}} \left( \frac{\ell^2 |k|}{r} \right) \,, \quad 
Z_- \equiv Z_{m\ell + \frac{1}{2}} \left( \frac{\ell^2 |k|}{r} \right) \,, \quad 
a_- \equiv \left(  \frac{i\, {\tilde \gamma} \cdot k}{|k|} \right) a_+\,,
\ee
so that 
\be
\phi_+ = r^{-2} Z_+ a_+ \,, \qquad \phi_- = r^{-2} Z_- a_- \,, 
\ee
and the final solution is
\be\label{phi0k}
\psi^{(0)} = e^{-i \om t \,+\, i k_1 x \,+\, i k_2 y}\,\, r^{-2} \left( \begin{array}{c} Z_+ a_+ \\ Z_- a_- \end{array}\right)\,.
\ee

Since we are interested in the IR regime of the backreacted metric which would affect the finite part of the holographic entanglement entropy, we will focus on the IR behavior of \eq{phi0k} for $r \to 0$, for which purpose the following asymptotic behavior of $H^{(1)}_{\n} (z) $ is employed:
 \be
 H^{(1)}_{\n} (z) \sim \sqrt{\frac{2}{\pi z}} \,e^{\,i\, ( z - \n {\pi \over 2} - {\pi \over 4} )}  \,,
 \quad\quad\quad{\mbox {for }} ~~ |z| \to \infty \,. 
 \ee
Again we choose the constant spinor $a_+ = \{0,~\a \}^{\mathrm{T}}$. The energy-momentum tensor ${\tilde {\bm \s}}_{\m \n}^{(0)}$ calculated through (\ref{sigma0}) takes the following form in the IR regime:
\be \label{sigma0k}
{\tilde {\bm \s}}_{\m \n }^{(0)IR}  ~\sim~  \left( \begin{array}{cccc} 
~{C_{00} / r^2}~ & ~~{C_{01} / r^2}~ & ~~{C_{02} / r^2}~ &~~{C_{03} / r^4}~\\  
{C_{01} / r^2} & {C_{11} / r^2} &{C_{12} / r^2} & {C_{13} / r^4}  \\  
{C_{02} / r^2}  &{C_{12} / r^2} & {C_{22} / r^2} & 0 \\
{C_{03} / r^4} & {C_{13} / r^4}  & 0 & {C_{33} / r^6}  
 \end{array}\right)   
\ee 
where $C_{\m \n}$ are functions of $\om, \,k_1, \,k_2$\,. If we ignore components other than the leading ${\cal {O}} (1 /r^6)$ one in \eq{sigma0k}, the backreacted metric would be similar to \eq{ansatz1}, i.e., diagonal and static. To match the ${\cal {O}} (1 / r^4)$ components in \eq{sigma0k}, $t$ and $x$ dependence need to be introduced in the metric. We take  the following ``minimally'' $t,\,x$-dependent metric ansatz 
\be \label{metric1k}
g_{\m \n }^{(1)IR}  ~\sim~  \left( \begin{array}{cccc} 
 - ({r^2 / \ell^2}) ( {b_t /  r^4}) & 0 & 0 & 0 \\  
0 &  ({r^2 / \ell^2}) ( {b_x  /  r^4})  & 0  & 0 \\  
0 & 0 &  ({r^2 / \ell^2}) ( {b_y   /  r^4}) & 0 \\
0  & 0 & 0 &  ({\ell^2 / r^2}) \left( {b_r  / r^4} \,+\, { c_r \,t / r^3 } \,+\, { d_r \,x / r^3 }\right)   
 \end{array}\right)  
\ee
in the sense that the leading terms of the L.H.S. of  \eq{einstein1} match $ r_L^2 \,{\tilde {\bm \s}}_{\m \n}^{(0)IR}$ up to ${\cal {O}} (1 / r^4)$  by choosing the coefficients $b_{\m}$, $c_r$ and $d_r$ properly. 
There is no $ {\cal O} (1 / r) $ behavior 
in \eq{metric1k}, indicating the vanishing of logarithmic term in the resulted holographic entanglement entropy hence the absence of fermi surface \cite{Ogawa:2011bz}. 
For simplicity we just focus on the $\om = k_1 = k_2 = 0$ case in the main text.


\section*{B. Holographic Entanglement Entropy  on a Strip  for the Deformed Fermion State}  

   The modular Hamiltonian is unknown for the strip case so that we cannot conclude the information inequality even we calculate the holographic entanglement entropy up to second order. However, it is still interesting to obtain it based on our backreacted metric, and compare with the results for the ball region. 
   
   Besides, one would like to see if there is logarithmic violation of the area law of entanglement entropy for the current case up to the second order. This kind of violation is expected if there is a Fermi surface. Naively, our fermion condensate may plays similar role of Fermi surface though the underlying CFT is strongly interacting and may not have quasiparticles for the existence of Fermi surface. Our calculation below indeed shows the null result. This is in contrast to the discussions in \cite{Ogawa:2011bz,Huijse:2011ef} by postulating some IR Lifschitz background with hyperscaling violation to induce logarithmic violation of area law.

\subsection*{B.1 The First Order}

We choose the region $A$ to be a strip of width $L_x$\,, i.e., $A:=\{(x,y)|-{L_x \over 2} \le x \le {L_x \over 2}, 0\le y\le L_y \}$ with $L_y$ very large.  Note that in the following we set the coupling constant $\eta_t=1$ for the bilinear coupling between axial fermion current and torsion. 

Consider the backreacted ansatz \eq{ansatz1} which takes the following form in the planar coordinates
 \be\label{ansatz1planar}
ds^2 = \frac{\ell^2 }{z^2} \left( - f(z) dt^2 + u(z) dx^2 + v(z) dy^2 + g(z) dz^2 \right)  \,,
\ee
where 
\bea
&& f(z) = 1 + \sk \,b_t \, {z^3 \over z_L^3}\,, \quad\quad\quad g(z) = 1 + \sk \,b_r \, {z^3 \over z_L^3} \,, \\
&& u(z) = 1 + \sk \,b_x \, {z^3 \over z_L^3}\,, \quad\quad\quad v(z) = 1 + \sk \,b_y \, {z^3 \over z_L^3} \,.
\eea

The area functional in this case is given by 
\be
\mbox{Area} = 2 \ell^2 L_y \int_{\epsilon}^{z_*} {dz \over z^2} \sqrt{[g(z) + u(z) x'(z)^2] v(z)}
\ee
where $\epsilon$ is the UV cutoff and $z_*$ is the turning point where $x'$ diverges. Varying the area functional we get the minimal surface $\gamma_A$ and thus the holographic  entanglement entropy:
\be\label{SA1}
S_A= {\ell^2 L_y \over 2G_N} \int_{\epsilon}^{z_*} {dz \over z^2}  
\sqrt{g(z) v(z) \over 1- {1 \over u(z)v(z)}{z^4\over z_*^4}}\;,
\ee
and $z_*$ is related to $L_x$ by the following relation
\bea\label{Lx1}
L_x &=& 2 \int_{\epsilon}^{z_*} dz\;  {z^2 \over z_*^2} \,\sqrt {{1 \over u(z)}} \sqrt{ {g(z) \over u(z) v(z)-{z^4\over z_*^4}} } \nn\\~\nn\\
&=&  2\int_{\epsilon}^{z_*} dz\;   {z^2 \over z_*^2} {1 \over \left(1-{z^4\over z_*^4}\right)^{1/2}} \nn \\
&& + \;\sk  \int_{\epsilon}^{z_*} dz\; \left[  (b_r - 2 b_x - b_y) {z^5 \over z_L^3 z_*^2} 
  + (b_x - b_r) {z^9 \over z_L^3 z_*^6} \right]  \frac{1}{\left(1-{z^4\over z_*^4}\right)^{3/2}}\;.
\eea
It turns out that the above integration is divergent at the order $\sk$, the reason is that the integrand becomes sharply divergent as $z \to z_*$\,. To cure this problem we replace the upper limit $z_*$ with $z_* - \d$ and let $\d \to 0$ after the integration, and find that the singular terms of $\d$ cancel with each other only if $b_x + b_y = 0$. Recall we set $b_x = b_y$ for homogeneity, we have \be\label{bxy0} b_x = b_y = 0\,. \ee
 Then \eq{Lx1} takes the form
\be \label{Lx12}
L_x ~\approx ~  c_0 z_* + \sk \,c_1 b_r  {z_*^4 \over z_L^3}  + \mathcal{O}(\sk^2) \,,
\ee
where $c_0 = 2\sqrt{\pi}\, \G(3/4) / \G(1/4)$, $c_1 = \pi /8$\,. From \eq{Lx12} we could solve $z_*$ in terms of $L_x$\,, and the holographic entanglement entropy \eq{SA1} up to order $\sk$ is worked out as the following:
\be
 S_A = S_A^{(0)} + \Delta S_A^{(1)}\,, 
\ee
where
\be
 S_A^{(0)} =  {\ell^2 L_y \over 2G_N \e} - d_0 \frac{\ell^2 L_y}{G_N L_x}\,, \quad \quad \quad
 \Delta S_A^{(1)} = \sk \,d_1 b_r {\ell^2  L_x^2 L_y \over G_N z_L^3}\,, 
\ee
where $d_0 = \pi \,\G(3/4)^2 / \G(1/4)^2$, $d_1 = \G(1/4)^2 / (128 \,\G(3/4)^2 )$. 

On the other hand, the change of energy could be obtained from \eq{stresstensor} as
\be
\Delta E_A^{(1)} = \int dx dy \,\Delta T_{tt} = \sk\, b_r {\ell^2 L_x L_y \over 8 \pi G_N z_L^3}   \,,
\ee
therefor we have the ``first law'' relation:
\be
\Delta E_A^{(1)}  = T_{ent} \cdot \Delta S_A^{(1)} \,,
\ee
with the entanglement temperature 
\be
T_{ent} = c \cdot L_x^{-1}\,,
\ee
where $c = 16 \,\G(3/4)^2 / (\pi \,\G(1/4)^2)$\,.  


\subsection*{B.2 The Second Order}

For the second order backreacted metric \eq{backreactg} and \eq{gzz}, we adopt the same prescription as in section \ref{sec 5}.  In this case the area functional for surface in constant time slice is given by
\be\label{areastrip}
\mbox{Area} = 2 \ell^2 L_y \int_{\epsilon}^{z_*} {dz \over z^2} \sqrt{G(z)+x'(z)^2} \,, 
\ee
where $\epsilon$ is the UV cutoff and $z_*$ is the turning point where $x'$ diverges. Varying the area functional we get the equation of motion for the minimal surface $\gamma_A$\,:
\be\label{xzprime}
x'(z) = {z^2 \over z_*^2} \sqrt{G(z) \over 1-{z^4\over z_*^4}}\,.
\ee
$z_*$ is related to $L_x$ by the relation
\be\label{Lx}
L_x=2 \int_{\epsilon}^{z_*} dz\;  x'(z)\;.
\ee

Setting $G(z)$ to be $G(z)^{(1)} = 1 + \sk\, b_r z^3 / z_L^3$ in equation \eq{xzprime}, we obtain the minimal surface  $x'(z)^{(1)}$ for the first order perturbed metric. The holographic entanglement entropy for stationary metric \eq{backreactg} and \eq{gzz} could be obtained up to order $\sk^2$ by substituting $x'(z)^{(1)}$ into \eq{areastrip} while setting $G(z)$ to be $G(z)^{(2)} = 1 + \sk\, b_r z^3 / z_L^3 + \sk^2 \, k_r z^6 / z_L^6$\;, the result is 
\be\label{SA2}
S_A ~=~  {\ell^2 L_y \over 2G_N \e} - j_0 \frac{\ell^2 L_y}{G_N L_x} + j_1 \mu_0 m \a \b \frac{L_x^2 L_y }{\ell^2}
+ j_2 (16 - 3 (7 \pi - 16) \mu_0^2 m^2 \ell^2 ) \a^2 \b^2 \frac{G_N  L_x^5 L_y}{\ell^8} + \mathcal{O}(G_N^2)\,,
\ee
where $j_i$'s are positive numerical constants with $j_0 = d_0$, $j_1 = 8 \pi d_1$\,.  Note that the condition for $S_2$ to be negative is different from the one given in \eq{finalB} for the ball region.


\vspace{1cm}








 
\vspace{1cm}
\begin{thebibliography}{99}



\bibitem{Ryu:2006bv} 
  S.~Ryu and T.~Takayanagi,
  ``Holographic derivation of entanglement entropy from AdS/CFT,''
  Phys.\ Rev.\ Lett.\  {\bf 96}, 181602 (2006)
  [hep-th/0603001].
  
 
 
\bibitem{Ryu:2006ef}
  S.~Ryu and T.~Takayanagi,
  ``Aspects of Holographic Entanglement Entropy,''
  JHEP {\bf 0608} (2006) 045
  [hep-th/0605073].
  
\bibitem{VanRaamsdonk:2016exw} 
  M.~Van Raamsdonk,
  ``Lectures on Gravity and Entanglement,''
  arXiv:1609.00026 [hep-th].

\bibitem{relativeE-1}
T. M. Cover, J. A. Thomas, ``Elements of Information Theory", 2nd Edition, ISBN: 978-0-471-24195-9. 
 
 
\bibitem{relativeE-2}
V. Vedral, ``The role of relative entropy in quantum information theory", Rev. Mod. Phys. 74, 197 (2002).

\bibitem{relativeE-3}
T. Sagawa, ``Second Law-Like Inequalities with Quantum Relative Entropy: An Introduction", arXiv:1202.0983 [cond-mat.stat-mech].

\bibitem{Casini:2008cr} 
  H.~Casini,
  ``Relative entropy and the Bekenstein bound,''
  Class.\ Quant.\ Grav.\  {\bf 25}, 205021 (2008)
  [arXiv:0804.2182 [hep-th]].

\bibitem{Wall:2010cj} 
  A.~C.~Wall,
  ``A Proof of the generalized second law for rapidly-evolving Rindler horizons,''
  Phys.\ Rev.\ D {\bf 82}, 124019 (2010)
  [arXiv:1007.1493 [gr-qc]].
  

\bibitem{Wall:2011hj} 
  A.~C.~Wall,
  ``A proof of the generalized second law for rapidly changing fields and arbitrary horizon slices,''
  Phys.\ Rev.\ D {\bf 85}, 104049 (2012)
  Erratum: [Phys.\ Rev.\ D {\bf 87}, no. 6, 069904 (2013)]
  [arXiv:1105.3445 [gr-qc]].

  
  
  
  
\bibitem{Bhattacharya:2012mi} 
  J.~Bhattacharya, M.~Nozaki, T.~Takayanagi and T.~Ugajin,
  ``Thermodynamical Property of Entanglement Entropy for Excited States,''
  Phys.\ Rev.\ Lett.\  {\bf 110}, no. 9, 091602 (2013)
  [arXiv:1212.1164].
  
  
  
\bibitem{Blanco:2013joa} 
  D.~D.~Blanco, H.~Casini, L.~Y.~Hung and R.~C.~Myers,
  ``Relative Entropy and Holography,''
  JHEP {\bf 1308}, 060 (2013)
  [arXiv:1305.3182 [hep-th]].
  
\bibitem{Wong:2013gua}
  G.~Wong, I.~Klich, L.~A.~Pando Zayas and D.~Vaman,
  ``Entanglement Temperature and Entanglement Entropy of Excited States,''
  JHEP {\bf 1312} (2013) 020
  [arXiv:1305.3291 [hep-th]].
  
  
\bibitem{Faulkner:2013ica} 
  T.~Faulkner, M.~Guica, T.~Hartman, R.~C.~Myers and M.~Van Raamsdonk,
  ``Gravitation from Entanglement in Holographic CFTs,''
  JHEP {\bf 1403}, 051 (2014)
  [arXiv:1312.7856 [hep-th]].
    
  
\bibitem{Nozaki:2013vta}
  M.~Nozaki, T.~Numasawa, A.~Prudenziati and T.~Takayanagi,
  ``Dynamics of Entanglement Entropy from Einstein Equation,''
  Phys.\ Rev.\ D {\bf 88} (2013) no.2,  026012
  [arXiv:1304.7100 [hep-th]].
  
\bibitem{Bhattacharya:2013bna}
  J.~Bhattacharya and T.~Takayanagi,
  ``Entropic Counterpart of Perturbative Einstein Equation,''
  JHEP {\bf 1310} (2013) 219
  [arXiv:1308.3792 [hep-th]].
  
\bibitem{Guo:2013aca} 
  W.~z.~Guo, S.~He and J.~Tao,
  ``Note on Entanglement Temperature for Low Thermal Excited States in Higher Derivative Gravity,''
  JHEP {\bf 1308}, 050 (2013)
  [arXiv:1305.2682 [hep-th]].

 

 
\bibitem{Swingle:2014uza} 
  B.~Swingle and M.~Van Raamsdonk,
 ``Universality of Gravity from Entanglement,''
  arXiv:1405.2933 [hep-th].
 
 
 
 
 
 
  
\bibitem{deBoer:2015kda} 
  J.~de Boer, M.~P.~Heller, R.~C.~Myers and Y.~Neiman,
  ``Holographic de Sitter Geometry from Entanglement in Conformal Field Theory,''
  Phys.\ Rev.\ Lett.\  {\bf 116}, no. 6, 061602 (2016)
  [arXiv:1509.00113 [hep-th]].
   
   
\bibitem{Czech:2016xec} 
  B.~Czech, L.~Lamprou, S.~McCandlish, B.~Mosk and J.~Sully,
  ``A Stereoscopic Look into the Bulk,''
  arXiv:1604.03110 [hep-th].
   
\bibitem{deBoer:2016pqk} 
  J.~de Boer, F.~M.~Haehl, M.~P.~Heller and R.~C.~Myers,
  ``Entanglement, Holography and Causal Diamonds,''
  arXiv:1606.03307 [hep-th].
 
 
 
\bibitem{Lin:2014hva} 
  J.~Lin, M.~Marcolli, H.~Ooguri and B.~Stoica,
  ``Locality of Gravitational Systems from Entanglement of Conformal Field Theories,''
  Phys.\ Rev.\ Lett.\  {\bf 114}, 221601 (2015)
  [arXiv:1412.1879 [hep-th]].
   
   
\bibitem{Lashkari:2014kda} 
  N.~Lashkari, C.~Rabideau, P.~Sabella-Garnier and M.~Van Raamsdonk,
  ``Inviolable energy conditions from entanglement inequalities,''
  JHEP {\bf 1506}, 067 (2015)
  [arXiv:1412.3514 [hep-th]].
   
   
\bibitem{Lashkari:2015hha} 
  N.~Lashkari and M.~Van Raamsdonk,
  ``Canonical Energy is Quantum Fisher Information,''
  JHEP {\bf 1604}, 153 (2016)
  [arXiv:1508.00897 [hep-th]].
   
   
\bibitem{Lashkari:2016idm} 
  N.~Lashkari, J.~Lin, H.~Ooguri, B.~Stoica and M.~Van Raamsdonk,
  ``Gravitational Positive Energy Theorems from Information Inequalities,''
  arXiv:1605.01075 [hep-th].
   
\bibitem{Jafferis:2015del} 
  D.~L.~Jafferis, A.~Lewkowycz, J.~Maldacena and S.~J.~Suh,
  JHEP {\bf 1606}, 004 (2016)
  doi:10.1007/JHEP06(2016)004
  [arXiv:1512.06431 [hep-th]].
   
\bibitem{Jafferis:2014lza} 
  D.~L.~Jafferis and S.~J.~Suh,
  ``The Gravity Duals of Modular Hamiltonians,''
  arXiv:1412.8465 [hep-th].
   


\bibitem{Dong:2016eik} 
  X.~Dong, D.~Harlow and A.~C.~Wall,
  ``Bulk Reconstruction in the Entanglement Wedge in AdS/CFT,''
  arXiv:1601.05416 [hep-th].
   
\bibitem{Almheiri:2014lwa} 
  A.~Almheiri, X.~Dong and D.~Harlow,
  ``Bulk Locality and Quantum Error Correction in AdS/CFT,''
  JHEP {\bf 1504}, 163 (2015)
  [arXiv:1411.7041 [hep-th]].
   
   
   
\bibitem{ElShowk:2012ht} 
  S.~El-Showk, M.~F.~Paulos, D.~Poland, S.~Rychkov, D.~Simmons-Duffin and A.~Vichi,
  Phys.\ Rev.\ D {\bf 86}, 025022 (2012)
  doi:10.1103/PhysRevD.86.025022
  [arXiv:1203.6064 [hep-th]].
   
 
\bibitem{Casini:2011kv} 
  H.~Casini, M.~Huerta and R.~C.~Myers,
  ``Towards a derivation of holographic entanglement entropy,''
  JHEP {\bf 1105}, 036 (2011)
  [arXiv:1102.0440 [hep-th]].
   


      
   
   
   
\bibitem{Breitenlohner:1982jf} 
  P.~Breitenlohner and D.~Z.~Freedman,
  ``Stability in Gauged Extended Supergravity,''
  Annals Phys.\  {\bf 144}, 249 (1982).
 ``Positive Energy in anti-De Sitter Backgrounds and Gauged Extended Supergravity,'' Phys.\ Lett.\ B {\bf 115}, 197 (1982).  
  
  
  

\bibitem{Myers:2010tj} 
  R.~C.~Myers and A.~Sinha,
  ``Holographic c-theorems in arbitrary dimensions,''
  JHEP {\bf 1101}, 125 (2011)
  [arXiv:1011.5819 [hep-th]].

\bibitem{Jafferis:2011zi} 
  D.~L.~Jafferis, I.~R.~Klebanov, S.~S.~Pufu and B.~R.~Safdi,
  ``Towards the F-Theorem: N=2 Field Theories on the Three-Sphere,''
  JHEP {\bf 1106}, 102 (2011)
  [arXiv:1103.1181 [hep-th]].

\bibitem{Ishihara:2012jg} 
  M.~Ishihara, F.~L.~Lin and B.~Ning,
  ``Refined Holographic Entanglement Entropy for the AdS Solitons and AdS black Holes,''
  Nucl.\ Phys.\ B {\bf 872}, 392 (2013)
  [arXiv:1203.6153 [hep-th]].

\bibitem{Rosenhaus:2014woa} 
  V.~Rosenhaus and M.~Smolkin,
  ``Entanglement Entropy: A Perturbative Calculation,''
  JHEP {\bf 1412}, 179 (2014)
  [arXiv:1403.3733 [hep-th]].
  
\bibitem{Rosenhaus:2014nha} 
  V.~Rosenhaus and M.~Smolkin,
  ``Entanglement Entropy Flow and the Ward Identity,''
  Phys.\ Rev.\ Lett.\  {\bf 113}, no. 26, 261602 (2014)
  [arXiv:1406.2716 [hep-th]].

\bibitem{Ben-Ami:2015zsa} 
  O.~Ben-Ami, D.~Carmi and M.~Smolkin,
  ``Renormalization group flow of entanglement entropy on spheres,''
  JHEP {\bf 1508}, 048 (2015)
  [arXiv:1504.00913 [hep-th]].
  
  
  
 

\bibitem{Faulkner:2014jva} 
  T.~Faulkner,
  ``Bulk Emergence and the RG Flow of Entanglement Entropy,''
  JHEP {\bf 1505}, 033 (2015)
  [arXiv:1412.5648 [hep-th]].
  

\bibitem{Vafa:2005ui} 
  C.~Vafa,
  ``The String landscape and the swampland,''
  hep-th/0509212.
  
  
\bibitem{Ooguri:2006in} 
  H.~Ooguri and C.~Vafa,
  ``On the Geometry of the String Landscape and the Swampland,''
  Nucl.\ Phys.\ B {\bf 766}, 21 (2007)
  [hep-th/0605264].
  








\bibitem{Eddington}
A. S. Eddington, "A generalisation of Weyl's theory of
the electromagnetic and gravitational fields, " Proc. R. Soc.
Lond. A 99, 104 (1921).
\\
A. S. Eddington, The Mathematical Theory of Relativity, 1924
2nd ed. (Cambridge University, Cambridge).


\bibitem{Cartan}
E. Cartan, "Sur une gdndralisation de la notion de
courbure de Riemann et les espaces a torsion, " C. R. Acad.
Sci. (Paris) 174, 593 (1922).




\bibitem{Hehl:1976kj} 
  F.~W.~Hehl, P.~Von Der Heyde, G.~D.~Kerlick and J.~M.~Nester,
  ``General Relativity with Spin and Torsion: Foundations and Prospects,''
  Rev.\ Mod.\ Phys.\  {\bf 48}, 393 (1976).
  


\bibitem{Shapiro:2001rz}
  I.~L.~Shapiro,
  ``Physical aspects of the space-time torsion,''
  Phys.\ Rept.\  {\bf 357} (2002) 113
  [hep-th/0103093].
  
\bibitem{Cai:2015emx}
  Y.~F.~Cai, S.~Capozziello, M.~De Laurentis and E.~N.~Saridakis,
  ``f(T) Teleparallel Gravity and Cosmology,''
  arXiv:1511.07586 [gr-qc].


  



\bibitem{Iqbal:2009fd}
  N.~Iqbal and H.~Liu,
  ``Real-time response in AdS/CFT with application to spinors,''
  Fortsch.\ Phys.\  {\bf 57} (2009) 367
  [arXiv:0903.2596 [hep-th]].
  
\bibitem{Liu:2009dm}
  H.~Liu, J.~McGreevy and D.~Vegh,
  ``Non-Fermi liquids from holography,''
  Phys.\ Rev.\ D {\bf 83} (2011) 065029
  [arXiv:0903.2477 [hep-th]].




\bibitem{Balasubramanian:1999re}
  V.~Balasubramanian and P.~Kraus,
  ``A Stress tensor for Anti-de Sitter gravity,''
  Commun.\ Math.\ Phys.\  {\bf 208} (1999) 413
  [hep-th/9902121].
  
\bibitem{deHaro:2000vlm}
  S.~de Haro, S.~N.~Solodukhin and K.~Skenderis,
  ``Holographic reconstruction of space-time and renormalization in the AdS / CFT correspondence,''
  Commun.\ Math.\ Phys.\  {\bf 217} (2001) 595
  [hep-th/0002230].
  
  
     

\bibitem{Ogawa:2011bz}
  N.~Ogawa, T.~Takayanagi and T.~Ugajin,
  ``Holographic Fermi Surfaces and Entanglement Entropy,''
  JHEP {\bf 1201} (2012) 125
  [arXiv:1111.1023 [hep-th]].
  

\bibitem{Huijse:2011ef}
  L.~Huijse, S.~Sachdev and B.~Swingle,
  ``Hidden Fermi surfaces in compressible states of gauge-gravity duality,''
  Phys.\ Rev.\ B {\bf 85} (2012) 035121
  [arXiv:1112.0573 [cond-mat.str-el]].
  
  
  
  
  
  
\bibitem{Swingle:2013zla} 
  B.~Swingle,
  J.\ Stat.\ Mech.\  {\bf 1410}, no. 10, P10041 (2014)
  doi:10.1088/1742-5468/2014/10/P10041
  [arXiv:1307.8117 [cond-mat.stat-mech]].
  
\bibitem{Cremonini:2013ipa} 
  S.~Cremonini and X.~Dong,
  Phys.\ Rev.\ D {\bf 89}, no. 6, 065041 (2014)
  doi:10.1103/PhysRevD.89.065041
  [arXiv:1311.3307 [hep-th]].





\bibitem{Nester} Private communication with J.~M.~Nester. 

\bibitem{Iyer:1994ys} 
  V.~Iyer and R.~M.~Wald,
  ``Some properties of Noether charge and a proposal for dynamical black hole entropy,''
  Phys.\ Rev.\ D {\bf 50}, 846 (1994)
  [gr-qc/9403028].
  
  
  
\bibitem{Hollands:2012sf} 
  S.~Hollands and R.~M.~Wald,
  ``Stability of Black Holes and Black Branes,''
  Commun.\ Math.\ Phys.\  {\bf 321}, 629 (2013)
  [arXiv:1201.0463 [gr-qc]].
  
  

\end{thebibliography}
 \end{document}